# Quadruple-well ferroelectricity and moderate switching barrier in defective wurtzite $\alpha$-Al$_2$S$_3$: a first-principles study

*Yuto Shimomura[1], Saneyuki Ohno[1,2], Katsuro Hayashi[1], Hirofumi Akamatsu[1]\**

[1]Department of Applied Chemistry, Graduate School of Engineering, Kyushu University, 744 Motooka, Nishi-ku, Fukuoka 819-0395, Japan.

[2]Institute of Multidisciplinary Research for Advanced Materials, Tohoku University, 2-1-1 Katahira, Aoba-ku, Sendai, Miyagi 980-8577, Japan.

**ABSTRACT**

 Wurtzite-type ferroelectrics are highly promising for next-generation microelectronic devices due to their ferroelectric properties and integration with exiting semiconductors. However, their high coercive fields, which are close to breakdown electric fields, need to be lowered. To deal with this issue and secure device reliability, much effort has been devoted to exploring novel wurtzite compounds with lower polarization switching barriers and implementing doping strategies. Here, we report first-principles calculations on polarization switching in cation-vacancy ordered wurtzite α-Al$_2$S$_3$, unveiling its uniaxial quadruple-well ferroelectricity and moderate switching barrier, 51 meV/cation, which is much lower than that of conventional wurtzite ferroelectrics. There are three important features relevant to the Al vacancies leading to the uncommon quadruple-well ferroelectricity and the moderate switching barrier: mitigation of cation-cation repulsion, structural flexibility that alleviates an in-plane lattice expansion, and formation of σ-like bonding states consisting of Al 3p$_z$ and S 3p$_z$ orbitals. Biaxial compressive strain and Ga doping lower the switching barriers by up to 40%. This study encourages experimental investigation of the ferroelectric properties for defective wurtzite α-Al$_2$S$_3$ as a new promising material with unconventional and intriguing ferroelectricity and suggests a potential strategy for reducing switching barriers in wurtzite ferroelectrics: *introducing cation vacancies*.

**1. Introduction**

Ferroelectric materials, characterized by electrically and/or mechanically switchable spontaneous polarization, are utilized in various devices such as piezoelectric actuators, pyroelectric sensors, and capacitors [1–3]. Recently, wurtzite-type compounds have been attracting much attention. This class of materials was known to be piezoelectric and pyroelectric owing to their polar crystal structures with *P*6$_3$*mc* space group symmetry. The possible ferroelectricity of wurtzite-type compounds is highly promising for cutting-edge computing and data storage devices due to their high spontaneous polarization. However, their ferroelectricity was not observed prior to dielectric breakdown because of their high coercive fields until Fichtner *et al*. clearly demonstrated ferroelectric switching for wurtzite-type Sc-doped AlN in 2019 [4]. Their work hinted at doping strategies that facilitate ferroelectric switching and ignited rapid progress in the exploration of wurtzite ferroelectrics. Consequently, ferroelectric switching has been reported for other doped wurtzite compounds such as $Al_{1-x}B_xN$, $Ga_{1-x}Sc_xN$, $Zn_xMg_{1-x}O$, and $Al_{1-x}Y_xN$ [5–8]. However, their high coercive fields, close to breakdown electric fields, reduce device reliability [9,10]. To address this issue, considerable effort has been made to explore novel wurtzite compounds with lower switching barriers and substantial breakdown electric fields [10–12] as well as to implement the doping strategy as mentioned above [4–8].

Wurtzite structures consist of an hcp anion arrangement with cations occupying half of the 4-fold coordinated tetrahedral sites (Figure 1b and d). The filled tetrahedra are all oriented upwards (or all oriented downwards), sharing a vertex. It is commonly considered that upon ferroelectric switching in typical wurtzite-type compounds such as pristine AlN, all the cations move collectively from the upwards-oriented tetrahedra to neighboring downwards-oriented tetrahedra along the *c*-axis direction by passing through anion basal triangles, which involves a bottleneck of switching [13–15]. The saddle point of the minimum energy pathway (MEP) is a nonpolar hexagonal-boron nitride (h-BN)-like structure with space group symmetry of $P6_3/mmc$, in which all the cations locate at 5-fold coordinated trigonal bipyramidal sites [14]. Doping isovalent cations and applying biaxial tensile strain reduce the switching barriers by destabilizing the polar wurtzite structures and/or stabilizing the saddle-point h-BN-like structures to achieve switching prior to breakdown [16,17]. Recently, Lee *et al.* reported a first-principles study suggesting that certain ternary wurtzite-type compounds containing two types of cations exhibit polarization switching not in a "collective" way via a h-BN-like structure, but in an "individual" or "stepwise" way via multiple local-minimum intermediate structures [18], highlighting that avoiding the saddle-point h-BN-like structures enables those multinary wurtzite compounds to have low switching barriers.

Here, we propose *introducing cation vacancies* as a potential strategy to lower the switching barriers through a first-principles study on the polarization switching in a defective wurtzite compound. γ-In$_2$Se$_3$ is known as a cation-vacancy ordered wurtzite (Figure 1a and c) [19,20]. About 30 compounds with γ-In$_2$Se$_3$-type structures are registered in the Inorganic Crystal Structure Database (ICSD). In this study, we focus on α-Al$_2$S$_3$, which consists of the most abundant elements found in the γ-In$_2$Se$_3$-type compounds. Our first-principles calculations in conjunction with solid-state nudged elastic band (SS-NEB) methods [21] unveiled MEPs for polarization switching in α-Al$_2$S$_3$, which manifest itself as uniaxial quadruple-well potential curves with four local-minimum states. The intermediate local-minimum structures encompass 5-fold trigonal bipyramidal coordination. The calculated switching barrier is much lower than that of conventional wurtzite ferroelectrics. Detailed analysis of the evolution of the atomic arrangements and chemical bonding during the polarization switching indicates that the Al vacancies play important roles in yielding the unconventional quadruple-well ferroelectricity and the moderate switching barrier. Our calculations also predict that biaxial compressive strain and Ga doping enable the tunability of switching barriers. The calculated piezoelectric constants suggest that piezoresponse force microscopy (PFM) allows us to distinguish the four local-minimum states with different polarization values experimentally. This work showcases introducing cation vacancies as a potential route to lower the switching barriers in wurtzite ferroelectrics as well as predicts the unconventional ferroelectricity in α-Al$_2$S$_3$.

## 2. Result and Discussion
### 2.1 Crystal Structure of α-Al$_2$S$_3$ and Polarization Switching Behavior

The crystal structures of cation-vacancy ordered wurtzite α-Al$_2$S$_3$ and conventional wurtzite ZnS are summarized in Figure 1. α-Al$_2$S$_3$ shows a polar and chiral $P6_1$ space group symmetry. Two-thirds of the tetrahedral sites in wurtzite-type structures are occupied by the Al atoms, while the remaining one-third are vacant [22,23]. In contrast to conventional wurtzite-type compounds, α-Al$_2$S$_3$ has two kinds of crystallographically inequivalent cation sites, both of which are located at Wyckoff positions 6a. These sites can be readily distinguished based on the position of the nearest-neighbor Al vacant sites, i.e., whether it locates either in the sulfur vertex or basal-triangle side along the $c$ axis. Figure 1e and f extract a column of the coordination tetrahedra along the $c$ axis from the whole structure of α-Al$_2$S$_3$ and ZnS, respectively. Hereafter, the tetrahedral sites with a nearest-neighbor vacant site in the vertex or basal-triangle side are referred to as T1 and T2, respectively. The suffixes "u" and "d" indicate the upwards-oriented or downwards-oriented tetrahedra, respectively.

The calculated lattice constants ($a$ = 6.41 Å, $c$ = 17.80 Å) for α-Al$_2$S$_3$ are in good agreement with the experimental ones ($a$ = 6.44 Å, $c$ = 17.90 Å) [23], substantiating that the GGA-PBEsol functional used in this study well reproduces crystal structures. The calculated electric polarization of α-Al$_2$S$_3$, 66 $\mu$C/cm$^2$, is comparable to or smaller than those of wurtzite ferroelectrics (e.g., 65 and 135 $\mu$C/cm$^2$ for ZnS and AlN, respectively) [13,15]. An MEP between the two oppositely polarized states with all the tetrahedra oriented upwards or downwards was calculated by using SS-NEB methods to elucidate the switching barrier and behavior (Figure 2a). Interestingly, α-Al$_2$S$_3$ has four local energy minima in the MEP, referred to as +HP, +LP, –LP, and –HP in the order of the magnitude of polarization. Here, the states with high, low, and zero absolute polarization values are denoted as HP, LP, and ZP, respectively, and the prefix "+" and "−" symbols indicate positive and negative polarization values, respectively. Figure 3 illustrates these five structures providing the local energy minima and the saddle point. Both +HP and +LP states possess $P6_1$ space group symmetry. The $P6_1$ symmetry is preserved during the switching pathway, except in the saddle-point ZP state, which has zero polarization and nonpolar $P6_122$ symmetry. The $6_1$ symmetry element is preserved during the entire polarization switching process, indicating that the polarization direction remains along the $c$ axis. In contrast to quadruple-well ferroelectrics such as $BiFeO_3$, which have multiple independent polarization axes [24], the four local-minimum states in the MEP of α-$Al_2S_3$ are polarized along the $c$ axis, revealing that α-$Al_2S_3$ is an unusual ferroelectric with a uniaxial quadruple-well potential as observed in $CuInP_2S_6$ [25].

Figure 2b illustrates the structural evolution of α-$Al_2S_3$ during the polarization switching as the columns of coordination tetrahedra along the $c$ axis for simplicity. All the six columns included in the unit cell are related to each other by the $6_1$ symmetry operation, which is preserved during the polarization switching. The atomic displacement in the column reflects the entire polarization switching process. In the initial high-polarization +HP state, the two inequivalent Al atoms, denoted as Al1 and Al2, occupy the T1u and T2u sites, respectively, enclosed by the upwards-oriented tetrahedra. Beyond a small potential hill, there emerges the +LP state with lower total energy and polarization (+30 $\mu C/cm^2$), where the Al1 atoms occupy the T1u sites whereas the Al2 atoms occupy 5-fold coordinated bipyramidal sites, referred to as B1 sites. In the structural evolution from the +HP to +LP states, the Al2 atoms move from the T2u sites towards the vacant sites and migrate to the bipyramidal B1 sites, whereas the Al1 atoms remain at the T1u sites. Upon transitioning from the +LP to ZP states, the Al2 atoms move from the bipyramidal B1 sites to the T1d sites in the downward-oriented tetrahedra, whereas the Al1 atoms still remain at the T1u sites, resulting in a half-switched state where half of the tetrahedra and the other half are oriented upward and downward with the polarization canceled out.

In the latter half of polarization switching from the ZP state to the –LP state to the final –HP state, the Al1 atoms at the T1u sites migrate to the T2d sites via the bipyramidal B2 sites. It should be noted that the Al1 (Al2) atoms sitting at the T1u (T2u) sites in the initial +HP state occupy the T2d (T1d) sites in the final –HP states, indicating the swapping of the crystallographically inequivalent sites during the polarization switching between the T1 and T2 sites. This highlights the nonconventional ferroelectricity of α-$Al_2S_3$.

Our first-principles phonon calculation revealed that both the HP and LP states were dynamically stable (Figure S1 and S2). Notably, $Al_2S_3$ with the LP structure has been synthesized by chemical vapor transport [23], while solid-state reaction methods yield α-$Al_2S_3$, which adopts the HP structure. These facts imply that these isosymmetric polymorphs are energetically antagonized at room temperature or above, although our first-principles calculations predict that the LP state is more stable than the HP state at 0 K.

**2.2 Comparison of Switching Behavior and Barriers in Wurtzite Compounds**

In this study, the switching barriers for quadruple-well potential surfaces are defined as the highest energy barrier between a valley and its neighboring peak towards the switching direction [18]. The switching barrier of α-$Al_2S_3$ corresponds to the total energy difference between the LP and ZP states. The switching barrier of α-$Al_2S_3$, 51 meV/cation, is approximately one-tenth that of AlN (523 meV/cation) and one-third that of $Al_{15/16}B_{1/16}N$ (150 meV/cation) [26]. The moderate switching barrier is anticipated to enable polarization switching prior to electric breakdown. We compare the switching behavior in α-$Al_2S_3$ and other wurtzite ferroelectrics to understand the moderate switching barrier in α-$Al_2S_3$ below.

In the binary wurtzite ferroelectrics such as pristine AlN, all the cations displace collectively from the upwards-oriented to the downwards-oriented tetrahedra during polarization switching [12,14]. In ternary wurtzite systems such as $Li_2SiO_3$, for which two-step polarization switching has been predicted by first-principles calculations, more electronegative atoms move first, followed by the migration of less electronegative atoms, with the first switching barrier being the highest [18]. In contrast to these cases, in α-$Al_2S_3$, consisting of only one type of cations, the Al atoms situated at the crystallographically different sites exhibit individual motion; the Al2 atoms exhibit displacement preceding that of the Al1 atoms, as shown in Figure 2b, predominantly because the Al2 atoms can displace towards the vacancy sites apart from the Al1 atoms, thereby mitigating cation-cation electrostatic repulsion.

In the saddle-point structures for MEPs in binary and ternary wurtzite-type compounds, cations locate at the trigonal bipyramidal sites [10,12,26]. The basal triangles of the bipyramids do not offer enough space for accommodating the cations, resulting in the expansion of the anion triangles. The resultant in-plane lattice expansion destabilizes the saddle-point structures. In sharp contrast, α-$Al_2S_3$ has the lowest total energy when half of the Al atoms locate at the bipyramidal sites (i.e., the LP states), highlighting the switching behavior quite different from that for other wurtzite ferroelectrics. Upon transitioning from the +HP to +LP states, the Al2 atoms move to the center of S triangles, accompanying an expansion of the triangle (See Figure S3). In conventional wurtzite ferroelectrics, such triangle expansion increases the in-plane lattice constants typically by 5% [14,16], destabilizing the h-BN-like saddle-point structures. In α-$Al_2S_3$, the in-plane lattice constant for the LP state is only 1% larger than that of the HP state (See Figure S4). Figure 3b and d depict the LP structure, in which the $AlS_5$ and $AlS_4$ polyhedra tilt with respect to the *c* axis, resulting in the buckling of close-packed sulfur basal planes, i.e., non-flat sulfur basal planes. The buckling alleviates the in-plane lattice expansion. In conventional wurtzite-type compounds, four anion tetrahedra are connected to each other via an S atom, "locking" the anion polyhedral network. Meanwhile, in α-$Al_2S_3$, just two or three tetrahedra are connected to each other due to the Al vacancies, leading to a flexibility of the polyhedral network.

Thus, in contrast to conventional wurtzite-type ferroelectrics, the cation vacancies mitigate the electrostatic repulsion between cations in α-$Al_2S_3$ (Figure 2b). Also, the disconnection of polyhedral network induced by the Al vacancies produces structural flexibility alleviating the in-plane lattice expansion. These features are considered as the primary reasons why the switching barrier of the defective wurtzite α-$Al_2S_3$ is much smaller than those of "filled" wurtzite-type compounds.

**2.3 Stabilization Mechanism of $AlS_5$ Trigonal Bipyramidal Coordination**

We have discussed above how the Al vacancies stabilize the intermediate structures including the LP states from the structural aspects. The half of Al atoms occupy the trigonal bipyramidal sites in the most stable intermediate LP structures for α-$Al_2S_3$, whereas, in binary and ternary wurtzite ferroelectrics, cations occupy the trigonal bipyramidal sites in the unstable saddle-point structures. It remains unclear why the LP states are stable in α-$Al_2S_3$ from the viewpoint of chemical bonding. To elucidate the underlying stabilization mechanism of the $AlS_5$ bipyramids in terms of chemical bonding, bond valence (BV) and crystal orbital Hamilton population (COHP) between the Al and S atoms were calculated. Here, BV and COHP describe the strength of chemical bonding based on the bond length and the interactions between atomic orbitals, respectively. The negative and positive values of COHP indicate bonding and antibonding interactions, respectively. We utilize negative COHP (–COHP) integrated with respect to energy (–ICOHP) within the valence bands, which indicates the magnitude of net energy gain due to bonding and anti-bonding interactions. Figure 4a labels the S atoms composing of AlS$_4$ tetrahedra or AlS$_5$ bipyramids as follows: the axial S atoms located at the +$c$ and –$c$ sides with respect to the Al atoms as ax1 and ax2, respectively, and the equatorial S atoms composing of basal triangles as eq1, eq2, and eq3. Figure 4b and c present the BVs and –ICOHP, respectively, for the Al1 and Al2 atoms against the S atoms for the ZP, +LP, and +HP states.

Transitioning from the +HP to +LP states, the Al2 atoms migrate from the tetrahedral T2u sites to the bipyramidal B1 sites (Figure 2b), as described above. In the +HP state, the BV sums and –ICOHP of both Al1 and Al2 atoms are primarily contributed to by four S atoms, ax1, eq1, eq2, and eq3, with minimal contribution from ax2, confirming 4-fold tetragonal coordination (Figure 4b and c). In contrast, in the +LP state, the BV sum and –ICOHP of Al2 atom are significantly contributed to by the five S atoms, corroborating 5-fold coordination rather than 3-fold coordination.

Let us pay attention to the chemical bonding evolution for the Al2 atoms, as it is crucial to understand the stabilization mechanism of the +LP state. Figure 4b reveals that the BV sum of the Al2 atoms is significantly smaller for the +LP state than for the +HP state, indicating that the ionic bonding between Al2 and S atoms become weaker upon transitioning from the +HP to +LP states. As shown in Figure 4c, however, the sum of –ICOHP for the Al2 atom is comparable in the +LP and +HP states, revealing that the energy gain due to covalent bonding remains upon the structural evolution. This motivates us to unravel the Al2-S covalent bonding that compensates for the poor ionic bonding in more detail.

Figure 4d and e depict the –COHP for the Al2-S bonds as a function of energy in the +LP and +HP states, respectively. A remarkable difference is seen just below the Fermi level; the axial and equatorial S atoms show bonding and antibonding contributions, respectively, in the energy range from –1 to 0 eV for the +LP state, yielding a net bonding contribution, whereas negligible –COHP is found in this energy range for the +HP state, indicating no bonding contribution. This remarkable difference is considered as a major factor stabilizing the LP states. Figure 4f and g illustrate the wavefunctions of eigenstates just below the Fermi level for the +LP and +HP states, respectively, both of which mainly consist of S $3p_z$ orbitals. In the eigenstate of the +LP state, the $3p_z$ orbitals of the ax2 S atoms elongate towards the Al2 atoms to overlap with the Al2 $3p_z$ orbitals, clearly indicating a σ-like bonding interaction between these orbitals. Meanwhile, the S $3p_z$ orbitals of the ax2 S atoms are localized and nonbonding in the eigenstate for the +HP state. The ax2 S atoms are 2-fold coordinated by Al atoms in the +HP state, rendering one of the 3p orbitals nonbonding. The Al2 displacements to the bipyramidal sites yield an additional Al coordination to the ax2 S atoms, causing the ax2 S $3p_z$ orbitals to take part in the bonding states with the Al $3p_z$ states. The formation of the bonding states in the LP states is considered as another factor stabilizing the LP state with respect to the HP state.

**2.4 Effects of Epitaxial Strain and Chemical Doping on Switching Barriers**

For wurtzite ferroelectrics, lowering the switching barrier is a key priority for wider practical applications. It has been demonstrated that epitaxial biaxial strain and chemical doping lower the switching barriers for ZnO and AlN by theory and experiments [16,17,27]. Here, the effects of biaxial strain and chemical doping on MEPs are investigated for $Al_2S_3$ to give insights into the experimental control of ferroelectric switching.

The reduction of switching barrier in α-$Al_2S_3$ requires the stabilization of the ZP states and/or the destabilization of the LP states. Without the constraint of strain, the in-plane lattice constant of the LP state is larger than that of the ZP state, as shown in Figure S4, implying that the MEP can be controlled by harnessing biaxial strain. Figure 5a shows calculated MEPs under biaxial strain. The total energies of the LP states become higher with respect to those of the ZP states under compressive (negative) biaxial strain, which is consistent with the in-plane lattice constant of the LP state larger than that of the ZP state (Figure S4). Figure 5b shows the switching barriers under strain, which correspond to the total energy difference between the LP and ZP states. The switching barriers decrease by 8.5 % with 2% compressive strain. The biaxial strain dependence of switching barriers for α-$Al_2S_3$ is contrary to that for wurtzite ferroelectrics such as ZnO and AlN, where the switching barriers are reduced by tensile biaxial strain [14]. This is because the in-plane lattice constant decreases in α-$Al_2S_3$ and increases in typical wurtzite ferroelectrics when climbing the potential hills for the polarization switching.

Next, we consider the chemical doping effects on the switching barriers. It has been reported that the switching barrier is successfully reduced for AlN by doping B atoms, which favor planar triangle coordination. It is not likely that B-atom doping helps lower the switching barrier for α-$Al_2S_3$, because it can stabilize the LP states with trigonal bipyramidal coordination. Here, we focus on doping of Ga atoms. Ga is considered as a prime candidate for the following three reasons; (1) Ga atoms are isovalent to Al. (2) α-$Ga_2S_3$ adopts γ-$In_2Se_3$ structure similarly to α-$Al_2S_3$ [28]. (3) The ionic radius of $Ga^{3+}$ is bigger than that of $Al^{3+}$. The cation-anion radius ratio of Al to S ($r_{Al}/r_S$ = 0.212) is less than 0.225, which is the minimum value for 4-fold tetrahedral coordination according to the Pauling's third rule. Meanwhile, $r_{Ga}/r_S$ is 0.255, indicating that Ga atoms favor tetrahedral coordination of S atoms [29]. It is expected from these facts that Ga-doping stabilize the ZP and HP states and destabilize the LP states, reducing the switching barrier.

Figure 5c shows the MEPs for representative structural models of $Al_{(12-x)/6}Ga_{x/6}S_3$ with $x = 0, 2, 4, 6, 8, 10, 12$. An increase in $x$ stabilizes the ZP and HP states with respect to the LP states, as expected. In the low-$x$ range ($0 \leq x \leq 6$), the pathway from the +LP to ZP to –LP states corresponds to the highest potential hill, whereas, in the high-$x$ range ($6 \leq x \leq 12$), the +HP-+LP pathway includes the highest potential hill. Figure 5d shows the box-and-whisker plot of the switching barriers against doping concentration $x$ in $Al_{(12-x)/6}Ga_{x/6}S_3$. The switching barrier decreases with an increase in $x$, and shows a minimum at $x = 6$, followed by an increase in the barrier above $x = 6$. The barrier is about 40% smaller for $x = 6$ compared to pristine $\alpha$-$Al_2S_3$. Thus, our calculations predict that Ga-doping facilitates the polarization switching for $\alpha$-$Al_2S_3$. Furthermore, Ga-doping modulates the relative energy relationship of the four different polarization states, which enables the control of stable phases.

**2.5 Piezoelectric Constants**

$\alpha$-$Al_2S_3$ was found to be a quite rare example of quadruple-well ferroelectrics. Its polarization is always oriented along the $c$ axis in the MEP with four local energy minima. The four polar states (±HP and ±LP) can be distinguished by PFM if they have distinct piezoelectric constants. Here, we calculated the piezoelectric constants for the +HP and +LP states. Table 1 summarizes the piezoelectric stress constants ($e_{33}$), elastic stiffness coefficients ($C_{33}$), and piezoelectric constants ($d_{33}$) for the +HP and +LP states. Their piezoelectric constants are comparable to that of pristine AlN (~5 pC/N) [30,31]. The piezoelectric constants of the LP and HP states are distinctly different so that these two states are distinguished for the $c$-plane cleaved single crystal samples by PFM, as in $CuInP_2S_6$ [25]. The four polar states can be detected by PFM since the positively and negatively polarized states show piezoresponse with opposite signs.

## 3. Conclusions

We unveiled an unusual ferroelectricity in a defective wurtzite α-$Al_2S_3$ using first-principles calculations. α-$Al_2S_3$ is predicted to be a rare example of quadruple-well ferroelectrics, which have four local energy minima in the MEPs. The intermediate lower-polarization states contain 5-fold coordinated $AlS_5$ trigonal bipyramids. The polarization switching barrier of α-$Al_2S_3$ (51 meV/cation, 1.0 meV/Å$^3$) is one order of magnitude smaller than that of a typical wurtzite ferroelectric AlN. The Al vacancies alleviate electrostatic repulsion between Al atoms and bring into structural flexibility mitigating elastic energy penalty during the polarization switching. The bonding interactions between Al and S $3p_z$ states play a role in stabilizing the $AlS_5$ bipyramidal coordination. Biaxial compressive strain and Ga-atom doping destabilize the intermediate lower-polarization structures, facilitating the polarization switching. In particular, 50% Ga substitution is predicted to reduce the switching barrier by about 40%. Our calculated piezoelectric constants revealed that PFM enables to distinguish the four polarized states. Overall, this study encourages the experimental investigation of an unconventional ferroelectric $Al_2S_3$ as a new ferroelectric material promising for computing and data storage devices. Notably, the predicted quadruple-well potential surface as well as its fine tunability with chemical doping enables innovative devices such as a multi-valued high-density memory. This work also provides a new strategy for reducing the switching barrier in wurtzite ferroelectrics: introducing cation defects.

## 4. Method

### 4.1 Density functional theory calculations

First-principles calculations were carried out based on density functional theory (DFT). We used the projector augmented-wave (PAW) method [32,33] and the GGA-PBEsol functional [34–36] as implemented in the Vienna Ab-initio Simulation Package (VASP 5.4.4) [37–40]. A plane-wave cutoff energy of 300 eV was used. The radial cutoffs of PAW data sets for Al, Ga, and S are of 1.4, 1.2, and 1.2 Å, respectively. Al 3s, 3p; Ga 3d, 4s, 4p; and S 3s, 3p states are treated as valence electrons. Γ-centered $3 \times 3 \times 2$ k-point mesh sampling was employed. The lattice constants and internal coordinates were optimized until residual stress and forces converged to 0.01 GPa and 1 meV/Å, respectively. Born effective charge tensors and piezoelectric stress tensors were obtained by using density functional perturbation theory (DFPT) calculations. For DFPT calculations, a plane-wave cutoff energy of 600 eV and Γ-centered $6 \times 6 \times 4$ k-point mesh sampling were used.

### 4.2 Minimum energy pathways

First, the higher symmetry structure of α-$Al_2S_3$ was searched using the spglib code [41]. The atomic displacements of the polar structure with respect to the higher symmetric structure were obtained using the STRUCTURE RELATIONS code in BCS [42–44]. By reversing the displacements, the polar structural models with polarization in the opposite direction were created. Structural relaxation was performed for the polar end structures. The intermediate images were generated by linear interpolation between the relaxed end structures. To represent a complicated pathway, we employed 32 intermediate images [11]. Solid-state nudged elastic band (SS-NEB) method [21] was employed to determine the switching pathways and barriers using VASP Transition State Theory (VTST) tools (VTST code-198) developed by Henkelman and Jonsson [45,46] except for the calculations under biaxial epitaxial strain. The polarization was calculated for each image using its structure and Born effective charges.

The effects of Ga doping on the switching barriers were also examined. Ga-doped structures were thoroughly searched using the CLUPAN code [47]. As mentioned above, the polarization switching in α-Al$_2$S$_3$ involves site exchange between the T1 and T2 sites. Depending on doped structural models, the initial structures are not equivalent to the final structures, leading to asymmetric MEPs. In this study, for simplicity, we chose the doped structural models for which the initial and final structures are equivalent to each other, and calculated MEPs by SS-NEB to obtain the switching barriers. We used the algorithm implemented in the pymatgen code [48] to check the consistence of the structures before and after the switching. Structural relaxation was performed for the obtained end structural models. After preparing initial pathways by the linear interpolation of the end structures, MEPs and switching barriers were calculated using SS-NEB methods. The number of data is 6 for $x = 2$ and 10; 15 for $x = 4$ and 8; and 14 for $x = 6$.

To examine the strain dependence of the switching barriers, MEPs were also calculated under the constraint of fixed in-plane lattice constants using NEB methods implemented in the VASP code. We defined in-plane biaxial stain as $s = (a - a_0)/a_0$, where $a_0$ is the in-plane lattice constant of the unstrained α-Al$_2$S$_3$ structure. The calculations were carried out within the $s$ range from –2 to 2%. Out-of-plane lattice constants and internal coordinates were optimized for the polar end structures of the switching pathways with fixed in-plane lattice constants. The initial pathways were created by the linear interpolation of the end structures.

**4.3 Piezoelectric constants**

In this study, we focused on the diagonal component of piezoelectric (strain) constant, $d_{33}$. The piezoelectric constants were derived from piezoelectric stress tensors and elastic constants according to the procedure described in Ref. [49]. We calculated elastic constants using the strain-energy relationship. In the Voigt notation, the elastic energy can be written as [50,51]

$$E = \frac{1}{2}\varepsilon_p C_{pq} \varepsilon_p, \quad (1)$$

where $C_{pq}$ is the elastic constants, and $\varepsilon_p$ is the strain. Consider the point group 6 for α-Al$_2$S$_3$, the elastic constant matrix $C$ is represented as [52]

$$\begin{pmatrix} C_{11} & C_{12} & C_{13} & 0 & 0 & 0 \\ C_{12} & C_{11} & C_{13} & 0 & 0 & 0 \\ C_{13} & C_{13} & C_{33} & 0 & 0 & 0 \\ 0 & 0 & 0 & C_{44} & 0 & 0 \\ 0 & 0 & 0 & 0 & C_{44} & 0 \\ 0 & 0 & 0 & 0 & 0 & \frac{1}{2}(C_{11}-C_{12}) \end{pmatrix} \quad (2).$$

The strain tensor $\varepsilon = (\varepsilon_1, \varepsilon_2, \varepsilon_3, \varepsilon_4, \varepsilon_5, \varepsilon_6)$ is defined as

$$\varepsilon = \begin{pmatrix} \varepsilon_1 & \frac{\varepsilon_6}{2} & \frac{\varepsilon_5}{2} \\ \frac{\varepsilon_6}{2} & \varepsilon_2 & \frac{\varepsilon_4}{2} \\ \frac{\varepsilon_5}{2} & \frac{\varepsilon_4}{2} & \varepsilon_3 \end{pmatrix} \quad (3).$$

By applying monoaxial strain $\varepsilon = (0, 0, \delta, 0, 0, 0)$ to the crystal, we obtained the elastic energy:

$$E = \frac{1}{2}C_{33}\delta^2 \quad (4).$$

Therefore, the elastic constants $C_{33}$ was determined from total energies versus strain along the $c$ axis.

**4.4 Others**

We used the LOBSTER code to perform COHP analysis [53–57]. Bond valence was calculated using the bond valence parameters reported by Brese and O'Keeffe [58]. Phonon band structures were calculated using the PHONOPY code [59,60]. Space group symmetry was determined using the spglib code [41]. The VESTA code was used to visualize crystal structures [61].

ASSOCIATED CONTENT

**Supporting Information**.

The Supporting Information is available in the latter part. It includes calculated phonon band structures of α-$Al_2S_3$ in the +HP and +LP states and changes in the area of the basal triangles involved with the Al1 and Al2 atoms and the lattice constants $a$ and $c$ during the polarization switching in α-$Al_2S_3$.

AUTHOR INFORMATION

**Corresponding Author**

*Hirofumi Akamatsu – Department of Applied Chemistry, Kyushu University, Motooka, Fukuoka 819-0395, Japan.

ACKNOWLEDGMENT

This research was supported by Japan Society of the Promotion of Science (JSPS) KAKENHI Grants Nos. JP17K19172, JP18H01892, JP19H00883, JP21K19027, JP21H05568, JP21H04619, JP23H02069, and JP23H01869. H.A. appreciates Murata Science Foundation and Collaborative Research Project of Laboratory for Materials and Structures, Institute of Innovative Research,

Tokyo Institute of Technology. S.O. gratefully acknowledges the Toyota Riken for financial support through a Rising Fellow Program. The computation was carried out using the computer resource offered under the category of General Projects by Research Institute for Information Technology, Kyushu University.

**Tables**

**Table 1.** Piezoelectric stress constants ($e_{33}$), elastic stiffness coefficients ($C_{33}$), and piezoelectric constants ($d_{33}$) for the +HP and +LP states of α-Al$_2$S$_3$.

| State | $e_{33}$ (C/m²) | $C_{33}$ (GPa) | $d_{33}$ (pC/N) |
|---|---|---|---|
| +HP | 0.654 | 90 | 7.3 |
| +LP | 0.246 | 65 | 3.8 |

**Figures**

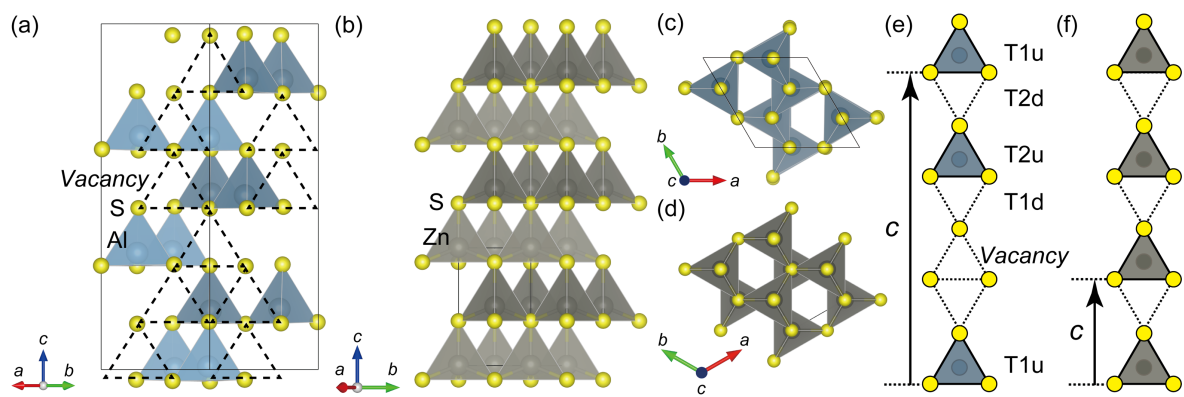

**Figure 1.** Crystal structures of **(a, c)** cation-vacancy ordered wurtzite α-$Al_2S_3$ with a γ-$In_2Se_3$-type structure (space group: $P6_1$) and **(b, d)** wurtzite ZnS (space group: $P6_3mc$). The solid lines indicate unit cells. The abbreviated illustrations of columns of coordination tetrahedra along the *c* axis for **(e)** α-$Al_2S_3$ and **(f)** ZnS.

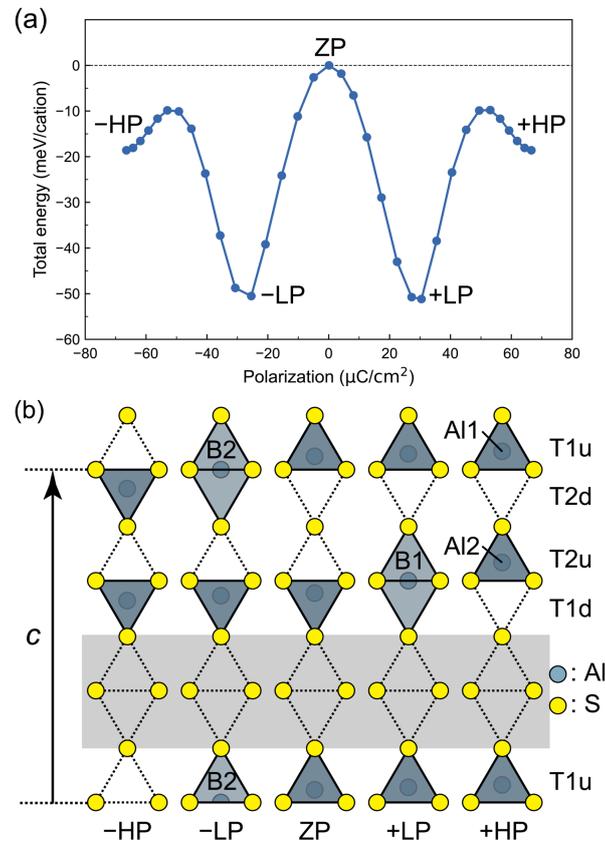

**Figure 2.** **(a)** Calculated minimum energy pathway associated with polarization switching in α-Al$_2$S$_3$. **(b)** Structural evolution during the polarization switching. For simplicity, the columns of coordination tetrahedra along the *c* axis are shown. The gray-colored region indicates the Al vacancy sites.

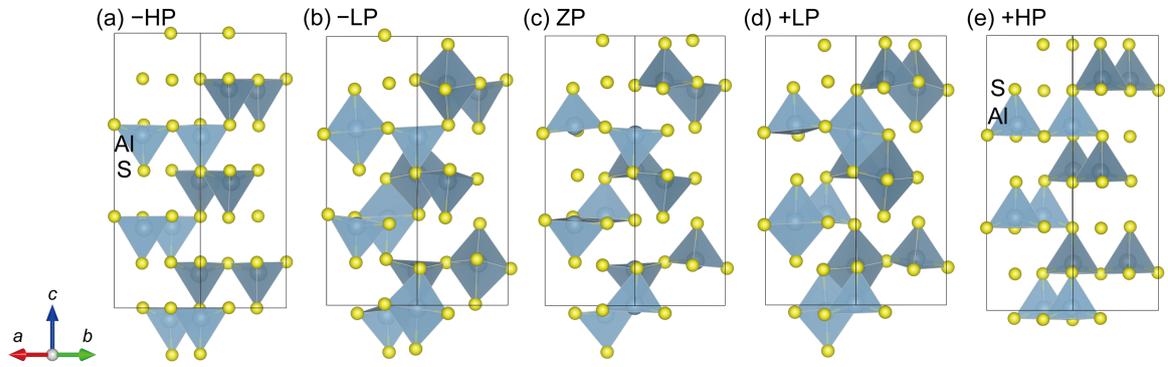

**Figure 3.** Schematic structures for (a) −HP, (b) −LP, (c) ZP, (d) +LP, and (e) +HP states.

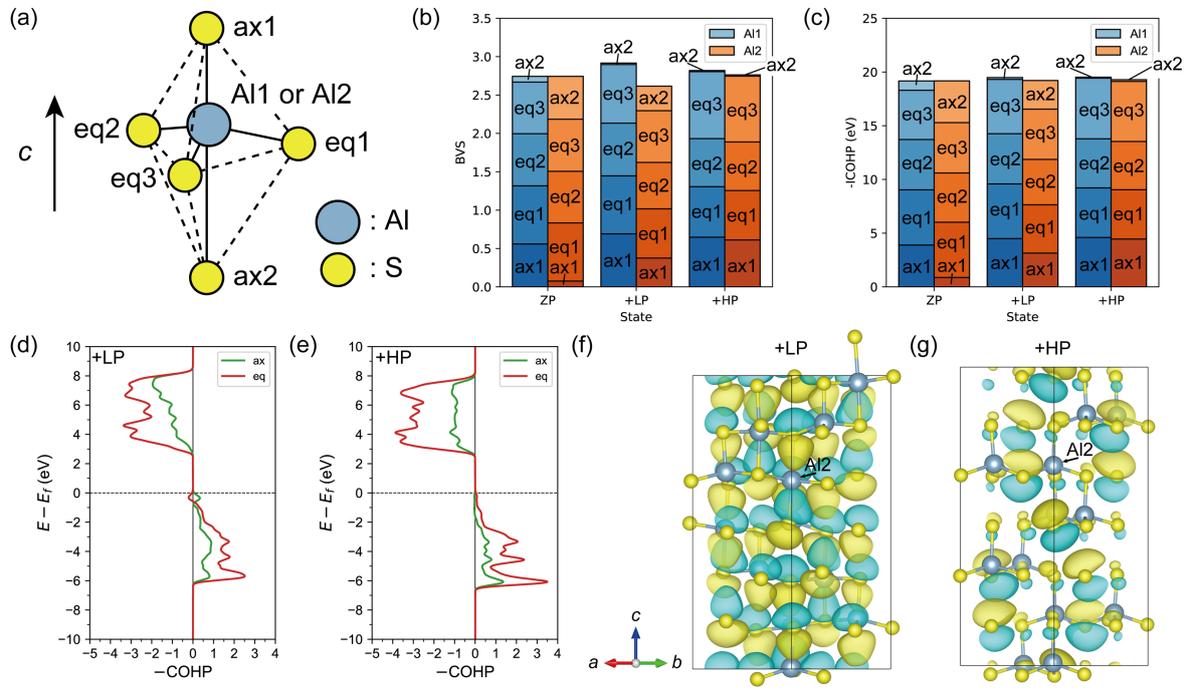

**Figure 4.** Chemical bonding in α-Al$_2$S$_3$. **(a)** Labeling of the S atoms surrounding the Al atom. **(b)** BV sums and **(c)** –ICOHPs between the Al and S atoms for the Al1 and Al2 atoms in the ZP, +LP, and +HP states of α-Al$_2$S$_3$. They are decomposed into the contributions from each bond. –COHPs plotted as a function of energy for the Al2-S bonds in the **(d)** +LP and **(e)** +HP states. They are decomposed into the contributions from axial and equatorial S atoms. Real part of the wave function for the bonding state just below the Fermi levels in the **(f)** +LP and **(g)** +HP states.

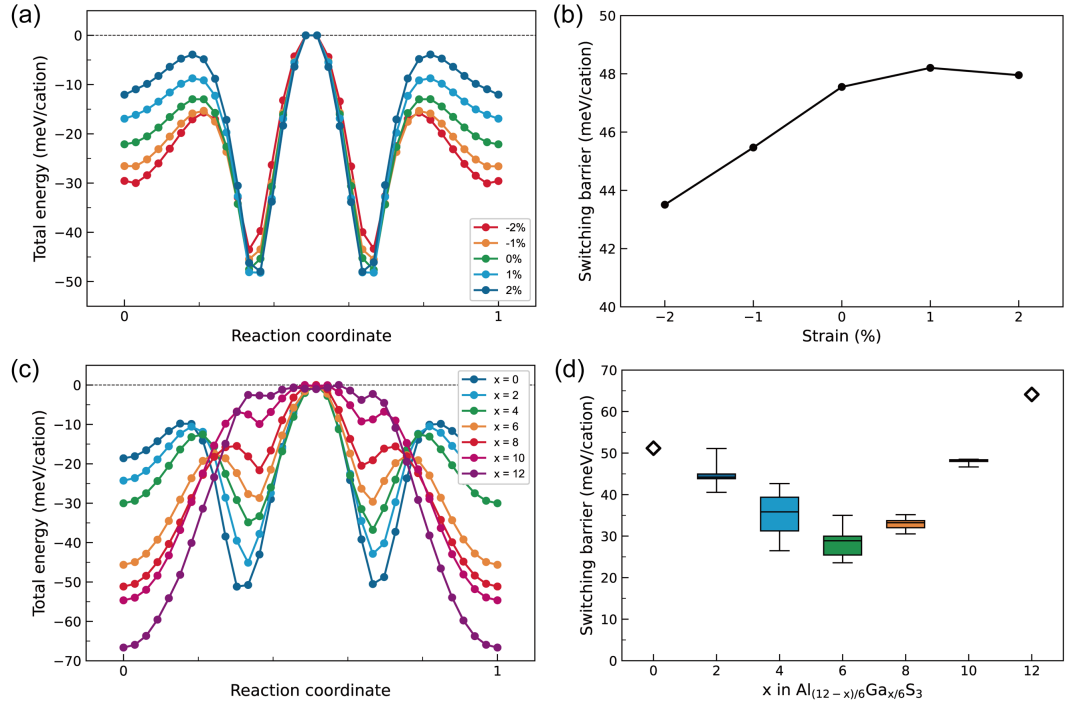

**Figure 5.** Switching barrier control by biaxial strain and chemical doping. **(a)** Total energy evolution in the MEPs of α-$Al_2S_3$ under biaxial strain ranging from –2 to 2 %. The highest total energies in each MEP are set to 0 meV/cation. **(b)** Switching barriers plotted against biaxial strain. **(c)** Total energy evolution in the MEPs for the representative structural models of $Al_{(12-x)/6}Ga_{x/6}S_3$ with $x$ = 0, 2, 4, 6, 8, 10, and 12. **(d)** Box plot of the switching barriers as a function of doping concentration $x$. The whiskers extend to the maximum and minimum data points.

# Supporting Information: Quadruple-well ferroelectricity and moderate switching barrier in defective wurtzite α-Al$_2$S$_3$: a first-principles study

*Yuto Shimomura[1], Saneyuki Ohno[1,2], Katsuro Hayashi[1], Hirofumi Akamatsu[1]\**

[1]Department of Applied Chemistry, Graduate School of Engineering, Kyushu University, 744 Motooka, Nishi-ku, Fukuoka 819-0395, Japan.

[2]Institute of Multidisciplinary Research for Advanced Materials, Tohoku University, 2-1-1 Katahira, Aoba-ku, Sendai, Miyagi 980-8577, Japan.

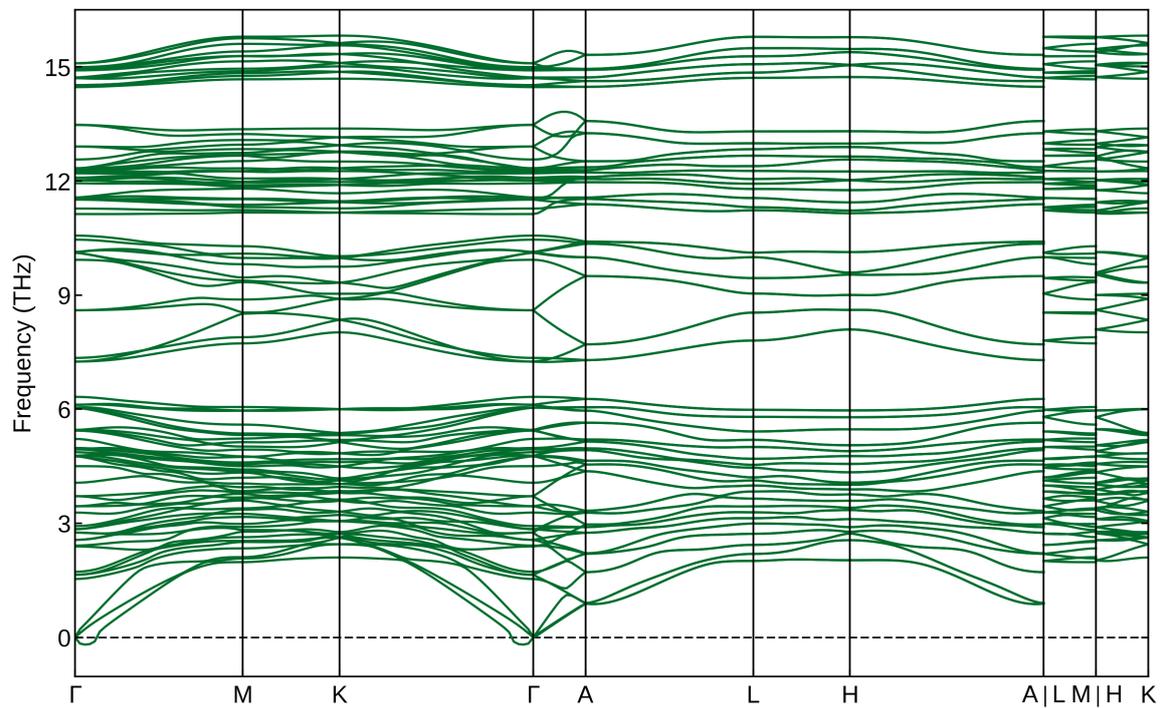

**Figure S1.** Phonon band structure of α-Al$_2$S$_3$ in the +HP states. The unstable branches near Γ point are considered to be interpolation artefacts since the size of supercell does not commensurate the wavevectors.

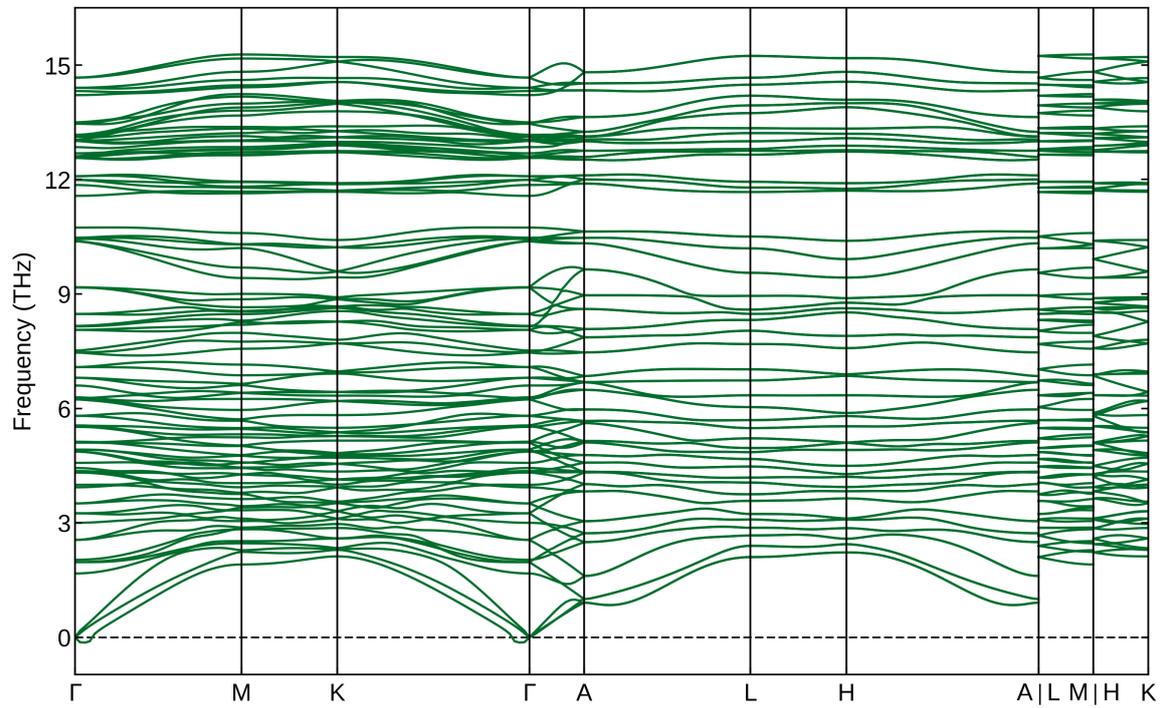

**Figure S2.** Phonon band structure of α-Al$_2$S$_3$ in the +LP states. The unstable branches near Γ point are considered to be interpolation artefacts since the size of supercell does not commensurate the wavevectors.

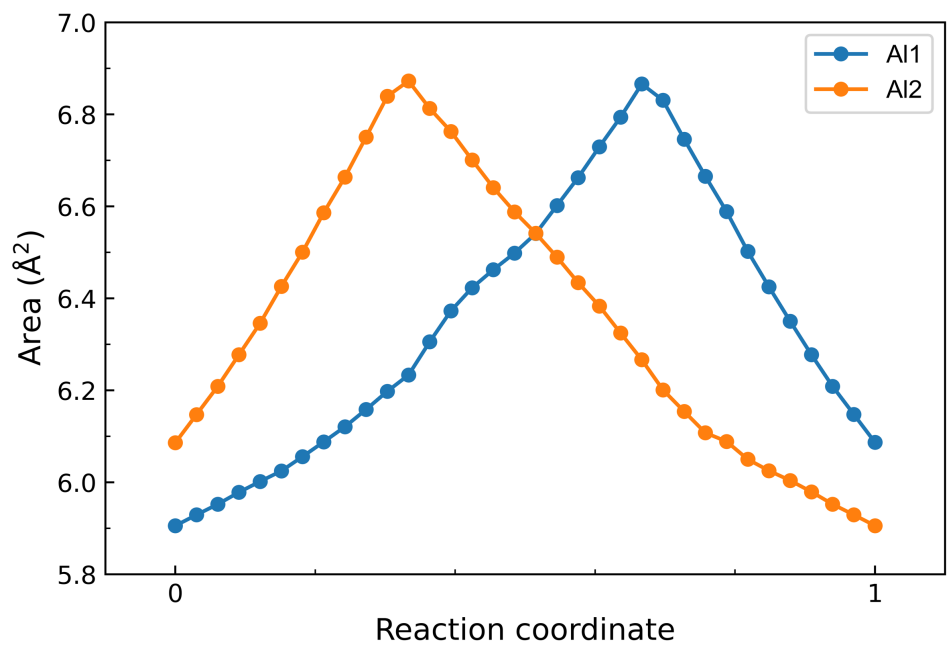

**Figure S3.** Change in the area of the basal triangles involved with the Al1 and Al2 atoms.

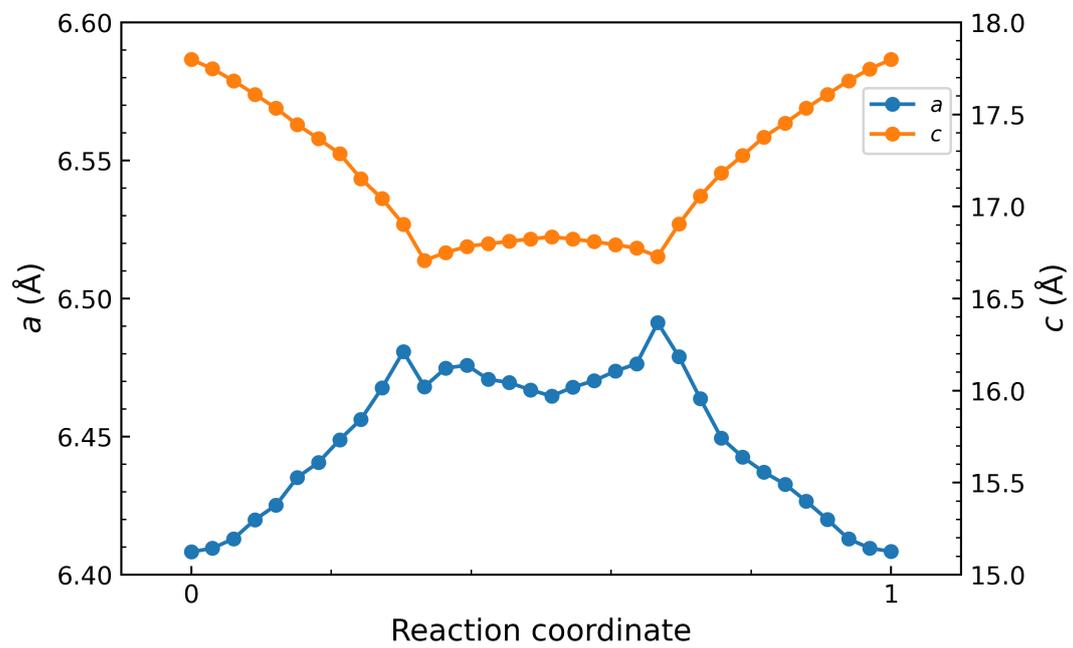

**Figure S4.** Variation of the lattice constants *a* and *c* during the polarization switching of α-Al$_2$S$_3$.